\documentstyle[12pt]{article}

\begin{document}
\title{A Bit-String Model for Biological Aging}
\author{T.J.P.Penna\footnote{Electronic address: gfitjpp@vmhpo.uff.br}\\
         Instituto de F\'{\i}sica, Universidade Federal
         Fluminense,\\
         Av. Litor\^anea s/n, 24210-340 Niter\'oi, RJ
         Brazil}
\date{\today}
\maketitle
\vfill
\begin{abstract}
We present a simple model for biological aging. We studied it
through computer simulations and we have found this model to reflect
some features of real populations.
\end{abstract}
\vfill
\centerline{\em to appear in Journal of Statistical Physics }
\newpage

The problem of aging has been studied recently \cite{Rose} also by
computer simulations, useful in understanding
how the survival rates for younger and older individuals evolve in
time and affect the preservation of the species
\cite{Stauffer}--\cite{Dasgupta}. The lowering of survival
rates as time goes by is called senescence. Mutations play an
important role in senescence modifying the survival rates either
increasing them (helpful mutations) or  decreasing them (deleterious
mutations). In this paper we introduce a simple model for aging
using the so-called ``bit string strategy''. This approach has been
applied to other biological systems\cite{biol}. Our model is
particularly suited for implementation in
computers, although analytical results may be obtained providing some
approximations. Since it is based on Boolean variables, bit-handling
techniques have been used~\cite{book}.  A complete description
of our code will be presented elsewhere~\cite{sttj}.

Let us consider a  population of $N(t)$ individuals, at the time $t$,
each one characterized by a genome which contains the information
when a mutation will occur in the lifetime of a given individual.  We
consider the time as a discrete variable ($t=1,2,\dots,B$) as suited
for implementation on computers. H\"otzel \cite{hotzel} simulated the
largest number of ages simulated up to now (five ages).
We denote each time step as one generation
since births can occur at each time step. Here, we will treat only
hereditary mutations, although somatic mutations can be incorporated
without great additional effort.  Hence, a genome is a
string of $B$ bits defined in the birth and kept unchanged during the
individual lifetime. The genome is built as follows: if an
individual has the $i$-th bit in genome set to one it will suffer a
deleterious mutation at age $i$. The parameter $T$ represents a
threshold, i.e., the maximum number of deleterious mutations that an
individual can suffer and stay alive. In order to include  the
effect of food and space restrictions and to keep the population
within the computer memory limits, we imposed the age-independent
Verhulst factor. Hence, an individual which has passed the threshold
test  only stays alive  with a probability $ ( 1-N (t )/N_{max} )$. The
next step is the birth: an individual whose age is larger than the
reproduction age $R$  generates one baby.  As far as we know,
this is the first model for biological aging where the reproduction
age appears as a parameter. Although sex is not a difficulty neither
for us nor for our model, we chose to work with asexual populations, for the
sake of simplicity -- sexual reproduction can be introduced, for
example,  by mixing the bit strings of two individuals (crossing
over). Thus, the baby's  genome will be the same as the parent,
except by a fraction $M$ of randomly changed bits (mutation rate,
hereafter).  As these new mutations are made randomly a baby can
suffer either a helpful mutation (flipping a bit set to one in the
parent's genome)  or a deleterious one (flipping a bit set to zero in
the parent's genome).  In summary, we have the following parameters
in this model: the genome size $B$, the threshold $T$, the maximum
number of individuals in the population $N_{max}$, the mutation rate $M$
and the minimum age at reproduction $R$.

After the presentation of the model we  are ready to show some
computer simulation results. First of all, it  is important to know
whether the proposed dynamics leads to stationary states. Fig.1 shows
results for the time evolution for
$N_{max}=10^5$, $T=2$, $R=8$ and $M=0,2$. The population first
decreases rapidly and grows again, as a signature of the Darwinistic
selection.  As  can be expected, the mutation rate controls the
relaxation time: the larger the mutation rate the lower the number of
generations to reach an equilibrium state. Fig.2 shows the
distribution by ages in the population for $T=4$, $M=1$ and $R=2,4,6$.
The aging can noted since the  frequency decreases stronger than exponentially
with  increasing age. This decrease  is more noticeable  for large ages than
the small ones. In fig.2 we
also can note that the larger the minimum age at
reproduction the larger the frequency of old ages. Old readers should
not celebrate this comment before checking the results presented in
fig.3. The average age at death decreases as soon as the age at
reproduction increases, consequently the total population decreases.
This behavior  has been found out in {\it Physella virgata virgata}
snail populations \cite{crowl}. There, the age of reproduction is controlled
by the presence of {\it Orconectes virilis} crayfish.  In our model
it is imposed as an additional parameter. It is worth stressing here
it is the first model -- to our knowledge -- where the maturity age
is introduced as a relevant parameter.

We can check the evolutive pressure in aging through the present
model, by measuring the frequency of deleterious mutation at each
age. Starting from an uniform distribution, fig.4 shows
the evolution of the frequency of bad mutations. The evolution
pressure is also larger for smaller minimum age at reproduction.
Note that the pressure is visible even at ages beyond the
minimum age. Therefore, the weak evolutive pressure on aging is also
reproducible in
this present model.

\bigskip
\noindent{\bf  Acknowledgments}

The author was introduced to this subject by D. Stauffer, an expert on aging.
The original code was written in C language and implemented in a personal
computer, a
lthough the FORTRAN translation does not
seem introduce any additional aging effect\cite{sttj}.

Financial support from the Brazilian Agencies CNPq and FINEP is
gratefully acknowledged.

\vspace{2cm}
{\Large\bf Figure Captions}

{\bf Fig. 1.} Number of individuals {\it vs.} generations for $M=0 (+)$ and
$M=2 (\bullet)$,
starting from $450000$ individuals. Mutations made the equilibrium state to be
reached faster.

{\bf Fig. 2.} Frequency of each age in population for $R=2 (\circ)$, $4 (+)$
and $8 (\Box)$.

{\bf Fig. 3.} Age at Death $(\Box)$ and Total Population $(\diamond)$ as
function of the minimum
age at reproduction $R$, for $M=2$ and $T=4$.

{\bf Fig. 4.} Evolution in time of the fraction of individuals suffering
mutations at ages:
$1 (\bullet)$, $6 (\Box)$, $12 (+)$ and $30 (\star)$. This data correspond to
$R=6$, $M=2$ and $T=4$ case.

\end{document}